\documentclass[11pt]{article}
\usepackage{amssymb,amscd,array}

\catcode `\@=11
\@addtoreset{equation}{section}

\newtheorem{thm}{Theorem}[section]

\def\qed{\blacksquare}
\newcommand{\be}{\begin{equation}}
\newcommand{\ee}{\end{equation}}
\newcommand{\bea}{\begin{eqnarray}}
\newcommand{\eea}{\end{eqnarray}}

\textwidth\def\kon#1#2{\vbox{\halign{##&&##\cr\lower4pt
\hbox{$\scriptscriptstyle\vert$}\hrulefill &\hrulefill\lower4pt
\hbox{$\scriptscriptstyle\vert$}\cr $#1$&$#2$\cr}}}

\def\al{\alpha}

\def\ro{\varrho}

\def\d{\partial}  
\def\=d{\,{\buildrel\rm def\over =}\,}

\def\sqr#1#2{{\vcenter{\vbox{\hrule height.#2pt\hbox{\vrule width.
#2pt height#1pt \kern#1pt \vrule width.#2pt}\hrule height.#2pt}}}}

\def\la{{\lambda}}

\begin{document}

\title{FROM MASSIVE GRAVITY TO MODIFIED GENERAL RELATIVITY II}
\author{D.R. Grigore 
\footnote{e-mail: grigore@theor1.theory.nipne.ro, grigore@theory.nipne.ro}
\\ Department of Theoretical Physics, 
\\Inst. for Physics and Nuclear
Engineering "Horia Hulubei"
\\ Institute of Atomic Physics
\\Bucharest-Magurele, P.O.Box MG6, ROMANIA
\\ G. Scharf
\footnote{e-mail: scharf@physik.unizh.ch}
\\ Institut f\"ur Theoretische Physik, 
\\ Universit\"at Z\"urich, 
\\ Winterthurerstr. 190 , CH-8057 Z\"urich, SWITZERLAND}

\date{}

\maketitle\vskip 3cm

\begin{abstract} 

We continue our investigation of massive gravity in the massless limit
of vanishing graviton mass. From gauge invariance we derive the most
general coupling between scalar matter and gravity. We get further
couplings beside the standard coupling to the energy-momentum tensor.
On the classical level this leads to a further modification of
general relativity.

\end{abstract}

\newpage

\section{Introduction}

In a previous paper \cite{Sch1} the massless limit of massive spin-2 quantum gauge
theory (called massive gravity for short) has been studied, and the
corresponding classical theory has been taken as an alternative to
general relativity. The reason why the limit of vanishing graviton mass
$m\to 0$ leads to a modification of general relativity is the vector
graviton field $v^\mu$ which is essential in the massive theory and
does not decouple from the symmetric tensor field $h^{\mu\nu}$ in the
massless limit. However, the theory studied in \cite{Sch1} is not yet complete
because the coupling to normal matter was described in the standard way
by means of the energy-momentum tensor of ordinary matter. Then there is no
direct coupling between the $v$-field and ordinary matter. But it is our
ultimate aim to derive all couplings from gauge invariance. For the scalar
matter couplings this is done in this paper. We will find that further
couplings between the scalar field $\Phi$ and the vector graviton field
$v^\mu$ with arbitrary coupling constant are possible in massive gravity.

As in \cite{Sch1} we then consider the massless limit $m\to 0$. In order to get a
non-trivial limit we have to choose the free coupling constants proportional
to the graviton mass $m$. This is not unusual because coupling terms with factors
$m$ appear also at other places in massive gravity. The surviving couplings
then lead to additional terms in the classical Lagrangean. This is a further
modification of general relativity.

The paper is organized as follows. In the next section we derive the most general gauge
invariant trilinear coupling between scalar matter and massive gravity. We apply
the descent method which was already used  for the construction of pure massive gravity
\cite{GS2}. To get uniqueness of the result the cohomological methods developed in
\cite{G2} and \cite{G3} have to be employed. We find five possible couplings where
three contain the vector graviton field $v^\mu$.

In Sect.3 we study second order gauge invariance which gives further restrictions on the
coupling. Only three coupling terms survive: one is the well-known coupling to the
energy-momentum tensor of the scalar field $\Phi$, the second is the $\Phi^3$ self-coupling
and there is one coupling to the $v$-field. However, the necessary finite renormalizations
generate new quartic couplings.

In the last section we investigate the new couplings in the limit of vanishing graviton
mass $m\to 0$. Since two of the quartic couplings contain $m$ in the denominator, a
non-trivial limit only exists if the (free) coupling constants are proportional to $m$.
The new coupling terms then give rise to the modification of general relativity mentioned
above.

\section{Gauge invariant couplings to scalar matter}

In \cite{GS1} we have analyzed the interaction of massless gravity with massive Yang-Mills fields and with scalar fields. The coupling of the free quantum fields can be obtained with the cohomology methods developed in \cite{G2} and \cite{G3}. The case of massive gravity can be analyzed with the same methods. 
We work in the same setting as in \cite{Sch1}, in particular we have the gauge structure
on the free asymptotic fields defined by the gauge charge operator $Q$ and the corresponding
gauge variation $d_Q$.
First we give the expression of the gauge invariant variables. It is convenient to introduce 
the following notations: 
\be
h \equiv \eta^{\mu\nu}h_{\mu\nu} \qquad
\hat{h}_{\mu\nu} \equiv h_{\mu\nu} - {1\over 2}~\eta_{\mu\nu}~h
\ee
and the we define the {\it Christoffel symbols} according to:
\be
\Gamma_{\mu;\nu\rho} \equiv \d_{\rho}\hat{h}_{\mu\nu} + \d_{\nu}\hat{h}_{\mu\rho} - \d_{\mu}\hat{h}_{\nu\rho}.
\ee
The expression
\be
R_{\mu\nu;\rho\sigma} \equiv \d_{\rho}\Gamma_{\mu;\nu\sigma} - (\rho \leftrightarrow \sigma)
\ee
is called the {\it Riemann tensor} and it is gauge invariant for massless and massive gravity also. In the case of massive gravity we have new gauge invariants namely the (symmetric) tensor

\be
\phi_{\mu\nu} \equiv 
- \d_{\mu}v_{\nu} - \d_{\nu}v_{\mu} + \eta_{\mu\nu} \d_{\rho}v^{\rho} + m~h_{\mu\nu}
\ee
and its trace:
\be
\phi \equiv \eta^{\mu\nu}~\phi_{\mu\nu}.
\ee
These expression are immediately proved to be gauge invariant. The same is true for their derivatives and the traceless part of these tensors. Let us denote by
$
R^{(0)}_{\mu\nu;\rho\sigma;\lambda_{1},\dots,\lambda_{n}},
\phi^{(0)}_{\mu\nu;\rho_{1}\dots\rho_{n}}, 
\phi^{(0)}_{;\rho_{1}\dots\rho_{n}} 
$ 
the traceless parts of these tensors. We denote the co-cycles of $d_Q$ by $Z_Q$. 
Then we have the following result \cite{G3}:
\begin{thm}
Let 
$
p \in Z_{Q}.
$
Then $p$ is cohomologous to a polynomial in the traceless variables described above.
\end{thm}
We note that in the case of null mass the operator 
$
d_{Q}
$
raises the canonical dimension by one unit and this fact is not true anymore in the massive case. We are lead to another cohomology group. Let us take as the space of co-chains the space 
$
{\cal P}^{(n)}
$
of polynomials of canonical dimension
$
\omega \leq n;
$
then 
$
Z_{Q}^{(n)} \subset {\cal P}^{(n)}
$
and
$
B_{Q}^{(n)} \equiv d_{Q}{\cal P}^{(n-1)}
$
are the co-cycles and the co-boundaries respectively. It is possible that a polynomial is a co-boundary as an element of
$
{\cal P}
$
but not as an element of
$
{\cal P}^{(n)}.
$
The situation is described by the following generalization of the preceding theorem.
\begin{thm}
Let 
$
p \in Z^{(n)}_{Q}.
$
Then $p$ is cohomologous to a polynomial of the form  
$
p_{1} + d_{Q}p_{2}
$
where
$
p_{1} \in {\cal P}_{0}
$ 
and
$
p_{2} \in {\cal P}^{(n)}.
$
\label{Q-cohomology}
\end{thm}

We will call the co-cycles of the type
$
p_{1}
$
(resp.
$
d_{Q}p_{2})
$
{\it primary} (resp. {\it secondary}). Using this result one can determine the most general form of the interaction between the massive gravity and a scalar field of mass $M$. We will call expressions of the type
$
d_{Q}B^{I} + i \d_{\mu}b^{I\mu}
$
{\it relative coboundaries}. 
\begin{thm}
Suppose that the interaction  Lagrangean $T$ between the massive gravity and a scalar field is trilinear in the fields (and their derivatives). Then $T$ it is relatively cohomologous to the following expression:
\bea
T = c_{1} \Phi \phi_{\mu\nu} \phi^{\mu\nu} + c_{2} \Phi \phi^{2}~ + c_{3} \Phi^{2}~\phi
+ c_4\left(\d_\mu\Phi\d_\nu\Phi h^{\mu\nu}-{1\over 2}M^2\Phi^2h\right) +c_5\Phi^3
\label{interaction}
\eea
\label{int}
i.e.
\be
d_{Q}T = i~\d_{\mu}T^{\mu}
\label{descent-tint}
\ee
with
\be
T^{\mu}_{0} = 
c_4 \left({1\over 2} u^{\mu} \d^{\nu}\Phi \d_{\nu}\Phi - u^{\nu} \d^{\mu}\Phi \d_{\nu}\Phi
- {1\over 2}M^{2} u^{\mu} \Phi^{2}\right). \label{2.7}
\ee
\end{thm}

Proof: (i) By hypothesis we have (\ref{descent-tint})
and the descent procedure (based on a variant of the Poincar\'e lemma \cite{G2}) leads to
\bea
d_{Q}T^{\mu} = i~\d_{\nu}T^{[\mu\nu]}.
\nonumber\\
d_{Q}T^{[\mu\nu]} = i~\d_{\rho}T^{[\mu\nu\rho]}
\nonumber \\
d_{Q}T^{[\mu\nu\rho]} = i~\d_{\sigma}T^{[\mu\nu\rho\sigma]}
\nonumber \\
d_{Q}T^{[\mu\nu\rho\sigma]} = 0
\label{descent-T-int}
\eea
where the carets indicate antisymmetry and can choose the expressions
$
T_{\rm int}^{I}
$
to be Lorentz covariant; we also have
\be
gh(T^{I}) = |I|, \omega(T^{I}) \leq 5. 
\ee 

>From the last relation in \ref{descent-T-int} we find, using the preceding Theorem 2.2, that
\be
T^{[\mu\nu\rho\sigma]} = d_{Q}B^{[\mu\nu\rho\sigma]} 
+ T_{0}^{[\mu\nu\rho\sigma]}
\ee
with
$
T_{0}^{[\mu\nu\rho\sigma]} \in {\cal P}_{0}^{(5)}
$
depending only on the invariants. It is easy to prove that such a (trilinear) expression does not exists so we have 
\be
T^{[\mu\nu\rho\sigma]} = d_{Q}B^{\mu\nu\rho\sigma}.
\ee
The third relation of the descent equations gives:
\be
d_{Q}(T^{[\mu\nu\rho]} - i~\d_{\sigma}B^{[\mu\nu\rho\sigma]}) = 0
\ee 
so we obtain again with the preceding Theorem
\be
T^{[\mu\nu\rho]} = B^{[\mu\nu\rho]} + i~\d_{\sigma}B^{[\mu\nu\rho\sigma]} 
+ T^{[\mu\nu\rho]}_0
\ee 
where
$
T_{0}^{[\mu\nu\rho]} \in {\cal P}_{0}^{(5)}
$
depends only on the invariants. Again, we can see that such an expression does not exists so we have
\be
T^{[\mu\nu\rho]} = B^{[\mu\nu\rho]} + i~d_{\sigma}B^{[\mu\nu\rho\sigma]}. 
\ee 
The second descent equation then gives 
\be
d_{Q}(T^{[\mu\nu]} - i~d_{\rho}B^{[\mu\nu\rho]}) = 0.
\label{tmunu}
\ee 

(ii) We obtain from the relation (\ref{tmunu}) with the preceding Theorem 2.2
\be
T^{[\mu\nu]} = d_{Q}B^{[\mu\nu]} + i~d_{\rho}B^{[\mu\nu\rho]} + T^{[\mu\nu]}_{0}
\ee 
where
$
T_{0}^{[\mu\nu]} \in {\cal P}_{0}^{(5)}.
$
The first descent equation gives the restriction:
\be
d_{Q}(T^{\mu} - \d_{\rho}B^{[\mu\nu]}) = \d_{\nu}T^{[\mu\nu]}_{0}
\ee 
so the divergence
$
\d_{\nu}T^{[\mu\nu]}_{0}
$
must be a coboundary. We do have a nontrivial expression for
$
T^{[\mu\nu]}_{0}
$
given by secondary cocycles. In the even sector with respect to parity we have
\bea
T^{[\mu\nu]}_{0} = g_{1} u^{\mu} u^{\nu} \Phi 
+ g_{2} u^{[\mu\rho]} u^{[\nu\sigma]} \eta_{\rho\sigma} \Phi
+ g_{3} u^{[\mu\nu]} u^{\rho} \d_{\rho}\Phi 
+ g_{4} (u^{[\mu\rho]} u^{\nu} -u^{[\nu\rho]} u^{\mu}) \d_{\rho}\Phi
\nonumber \\
+ g_{5} (u^{[\mu\rho]} \d^{\nu}\Phi - u^{[\nu\rho]} \d^{\mu}\Phi) u_{\rho} 
+ g_{6} (u^{\mu} u_{\rho} \d^{\nu}\d^{\rho} - u^{\nu} u_{\rho} \d^{\mu}\d^{\rho})
\eea
and in the odd sector we have the expression 
$
\epsilon^{\mu\nu\rho\sigma} T^{\prime}_{[\rho\sigma]}
$
where 
$
T^{\prime}_{[\rho\sigma]}
$
has the same form as above but with
$
g_{j} \rightarrow g_{j}^{\prime}.
$
Here we have used the following notation:
\be
u^{[\mu\nu]} = \d^{\mu}u^{\nu} - \d^{\nu}u^{\mu}
\ee
One computes the divergence
$
\d_{\nu}T^{[\mu\nu]}_{0}
$
and requires that it is a coboundary. After some computations one finds out that the remaining terms can be grouped into a relative coboundary i.e.
$
T^{[\mu\nu]}_{0} = d_{Q}b^{\mu\nu} - i \d_{\rho}b^{[\mu\nu\rho]}.
$
It follows that we have
\be
T^{[\mu\nu]} = d_{Q}B^{[\mu\nu]} + i~d_{\rho}B^{[\mu\nu\rho]} 
\ee 
if we redefine the expressions
$
B^{[\mu\nu]}
$
and
$
B^{[\mu\nu\rho]}.
$

The first descent equation gives
\be
d_{Q}(T^{\mu} - i \d_{\rho}B^{[\mu\nu]}) = 0
\ee 
so if we use the Theorem 2.2 we find
\be
T^{\mu} = d_{Q}B^{\mu} + i \d_{\rho}B^{[\mu\nu]} + T^{\mu}_{0}
\ee
where
$
T^{\mu}_{0} \in {\cal P}_{0}^{(5)}.
$
If we substitute this in the starting relation (\ref{descent-tint}) we get the consistency condition
\be
d_{Q}( T^{\mu} - i \d_{\mu}B^{\mu}) = i \d_{\mu}T^{\mu}_{0}
\ee
i.e. the divergence
$
\d_{\mu}T^{\mu}_{0}
$
must be a coboundary. The generic form of 
$
T^{\mu}_{0}
$
is again a secondary cocycle. In the even sector with respect to parity we have:
\be
T^{\mu}_{0} = f_{1} u^{\mu} \Phi^{2} + f_{2} u^{\mu} \d^{\nu}\Phi \d_{\nu}\Phi
+ f_{3} u^{\nu} \d^{\mu}\Phi \d_{\nu}\Phi + f_{4} u^{\mu\nu} \Phi \d_{\nu}\Phi
+ f_{5} u_{\nu} \Phi \d^{\mu}\d^{\nu}\Phi.
\ee

In the odd sector we have
\be
T^{\mu}_{0} =  f^{\prime} \epsilon^{\mu\nu\rho\sigma} u_{\nu\rho} \Phi \d_{\sigma}\Phi. 
\ee
We compute the divergence
$
\d_{\mu}T^{\mu}_{0}
$
and the consistency condition leads to
\be
T^{\mu}_{0} = 
f \left({1\over 2} u^{\mu} \d^{\nu}\Phi \d_{\nu}\Phi - u^{\nu} \d^{\mu}\Phi \d_{\nu}\Phi
- {1\over 2}M^{2} u^{\mu} \Phi^{2}\right) + d_{Q}b^{\mu}_{0} + \d_{\nu}b^{\mu\nu}_{0}\label{2.26}
\ee
for some arbitrary constant $f$. One can get rid of the relative coboundary by redefining the expressions
$B^{\mu}$
and
$
B^{\mu\nu}.
$
Moreover one proves that
$
\d_{\mu}T^{\mu}_{0} = - i d_{Q}t
$
where
\be
t \equiv f \left(h_{\mu\nu} \d^{\mu}\Phi \d^{\nu}\Phi -{1\over 2} M^{2} h \Phi^{2}\right) 
\ee

The starting relation (\ref{descent-tint}) is now
\be
d_{Q}( T - t - i \d_{\mu}B^{\mu} ) = 0
\ee
so that a final use of the Theorem 2.2 gives
\be
T = t + d_{Q}B + i \d_{\mu}B^{\mu} + T_{0}
\ee
with
$
T_{0} \in {\cal P}_{0}^{(5)}.
$
The generic form of
$
T_{0}
$
is
\be
T_{0} = c_{1} \Phi \phi^{(0)}_{\mu\nu} \phi^{(0)\mu\nu} + c_{2} \Phi \phi^{2}~ 
+ c_{3} \Phi^{2}~\phi + c_{4}\Phi^3
\ee
The expression from the statement follows easily: we can replace
$
\phi^{(0)}_{\mu\nu}
$
by
$
\phi_{\mu\nu}
$
if we redefine the constant
$
c_{2}
$
and $T^\mu$ follows from (\ref{2.26}).
$\qed$

\section{Second order gauge invariance}

In second order we must construct chronological products 
$
T(x,y)
$ 
and
$T_\mu (x,y)$ such that
\be
d_QT(x,y)=i{\d\over \d x^\mu}T^\mu (x,y)+x\leftrightarrow y\label{3.1}
\ee
is verified. The construction procedure is well-known: one first computes
the causal commutators 
$
[T(x),T(y)]
$ 
and 
$
[T_\mu (x),T(y)]
$ 
and substitutes the causal Pauli-Jordan distributions in the tree graph contributions by
Feynman propagators 
$
D^F(x-y)
$. 
If on the right-hand side of (\ref{3.1}) a wave operator 
$
\d^2
$ 
operates on 
$
D^F
$ 
we obtain a local term 
$
\sim\delta (x-y)
$.
These anomalies must be compensated by finite renormalizations.

The generic form of the anomaly is
\be
A(x,y)=\delta(x-y)a(x)+[\d_\mu^x\delta(x-y)]a^\mu(x,y)\label{3.2}
\ee
The total anomaly is obtained by adding the contribution $A(y,x)$ with $x$, $y$
interchanged. Then the terms with $\d\delta$ can be combined by means of the
identity
\be
[\d_\mu^x \delta(x - y)]f(x,y)+x\leftrightarrow y=[\d_{\mu}^yf-\d_\mu^xf]\delta(y-x)],
\label{3.3}
\ee
which follows by smearing with symmetric test functions; this is the right
test function space here, due to the symmetry of the chronological products.
Then the total anomaly is equal to
\be
A_{\rm tot}(x,y)=[2a(x)+\d_\mu^y a^\mu-\d_\mu^x a^\mu]\delta(x-y)
\equiv A(x)\delta(x-y).\label{3.4}
\ee

The cancellation of the anomalies is equivalent to 
\be
A_{\rm tot}(x,y) = d_{Q}R(x,y) - i \d_{\mu} R^{\mu}(x,y)+x\leftrightarrow y;\label{3.5}
\ee
here the expressions 
$
R(x,y)
$ 
and 
$
R^{\mu}(x,y)
$ 
are {\it finite renormalizations}: these are quasilocal operators: 
\be
R(x,y) = \delta(x-y) B(x) + \cdots\label{3.6}
\ee
and
\be
R^{\mu}(x,y) = \delta(x-y) B^{\mu}(x) + \cdots\label{3.7}
\ee
where $B$ and 
$
B^{\mu}
$ 
are some Wick polynomials and $\cdots$ are similar terms with derivatives on the delta distribution. Indeed, in this case one can eliminate the anomaly by redefinition of the chronological products
\be
T(x,y) \rightarrow T(x,y) + R(x,y)\label{3.8}
\ee
and
\be
T^{\mu}(x,y) \rightarrow T^{\mu}(x,y) + R^{\mu}(x,y).\label{3.9}
\ee

One can prove that the cancellation (\ref{3.5}) of the anomalies is achieved if we can write 
the operator part $A(x)$ in (\ref{3.4}) in the form
\be
A = d_{Q}B - i \d_{\mu}B^{\mu}.\label{3.10}
\ee
In fact, the derivative terms in (\ref{3.2}) can be combined with help of the identity
\be
\d_\mu^x[B^\mu(x,y)\delta(x-y)]+x\leftrightarrow y=[\d_\mu^x B^\mu+\d_\mu^y B^\mu]
\delta(x-y).\label{3.11}
\ee

The terms in 
$
T_\mu
$ 
which generate anomalies are the following ones:
\bea
T_\mu^{an}=u^\al(2\d_\al h^{\ro\nu}\d_\mu h_{\ro\nu}-\d_\al h\d_\mu h
+2\d_\al u^\nu\d_\mu\tilde u_\nu-2\d_\al\d_\nu u^\nu\tilde u_\mu)
\nonumber \\
+2\d_\nu u^\nu h^{\al\ro}\d_\mu h_{\al\ro}-\d_\nu u^\nu h\d_\mu h+
2\d_\nu u_\al h^{\al\nu}\d_\mu h-4\d^\nu u_\al\d_\mu h^{\al\ro}h_{\nu\ro}
\nonumber \\
-4u^\al\d_\al v^\nu\d_\mu v_\nu 
-c_4u^\al\d_\al\Phi\d_\mu\Phi.
\eea
Here we have put the gravitational coupling constant $\kappa=1$ for simplicity.
According to theorem \ref{int} the first order coupling to the scalar field 
$\Phi$
of mass $M$ is given by
\bea
T_\Phi=c_1\Phi\phi_{\mu\nu}\phi^{\mu\nu}
+c_2\Phi(mh+2\d_\mu v^\mu)^2
+c_3\Phi^2(mh+2\d_\mu v^\mu)
\nonumber \\
+c_4\left(\d_\mu\Phi\d_\nu\Phi h^{\mu\nu}-{1\over 2}M^2\Phi^2h\right)+c_5\Phi^3.\label{3.13}
\eea
We first consider the couplings linear in $\Phi$, i.e. with coefficients
$c_1, c_2$.

\begin{thm}
Second order gauge invariance implies $c_1=0$ and $c_2=0$.
\end{thm}

{\bf Proof:} To prove this result it is sufficient to find anomalies
with 
$c_1$ 
or 
$c_2$, 
which cannot be compensated. For 
$c_1$ 
we consider the commutator
\bea
-8c_1u^\la \d_\la v^\nu[\d_\mu v_\nu(x),\phi^{\al\beta}(y)]\phi_{\al\beta}(y)\Phi
\nonumber \\
\eea
As described above the commutator gives a causal propagator which in the chronological
product becomes a Feynman propagator. Applying the derivative 
$
\d/\d x^\mu
$ 
we get a 
$
\d^2 D^F
$ 
leading to the anomaly 
\be
A_1=4ic_1u^\la\d_\la v^\nu(x)\Phi(y)\Bigl(2\phi_{\al\nu}(y)\d_y^\al-\phi(y)\d_\nu^y\Bigl)
\delta(x-y).\label{3.14}
\ee

In the same way we consider the commutator
\be
-8c_2u^\al\d_\al v^\nu[\d_\mu v_\nu(x),\phi(y)]\phi(y)\Phi(y).
\ee
Here the resulting anomaly is equal to
\be
A_2=-8ic_2u^\al\d_\al v^\nu(x)\Phi(y)\phi(y) \d^y_\nu\delta(x-y).
\ee
There are no other anomalies with Wick monomials $uv\phi\Phi$, $uv\phi_{\mu\nu}\Phi$,
respectively. Consequently, $A_1$ and $A_2$ must cancel against each other. For the
last Wick monomial we see from (\ref{3.14}) that $c_1$ must be 0 and hence, $c_2$
must also vanish.

The situation is non-trivial for the remaining couplings which are bilinear in 
$\Phi$.
\begin{thm}
Second order gauge invariance implies 
$c_4=-2$, 
but
$c_3$ 
and 
$c_5$ 
remain unrestricted. In the second-order chronological products the following finite renormalizations are necessary
\be
T(x,y)=T^F(x,y)+i\delta(x-y)N(x)\qquad T_\mu(x,y)=T_\mu^F(x,y)+i\delta(x-y)N^\mu(x)\label{3.16}
\ee
where
\bea
N=2\Phi^2\Bigl\{M^2(2h^{\mu\nu}h_{\mu\nu}-h^2)+c_3\Bigl[m(2h^{\mu\nu}h_{\mu\nu}
-h^2)+{8\over m}(\d_\mu v^\mu\d_\nu v^\nu-\d_\mu v^\nu\d_\nu v^\mu)
\Bigl]
\nonumber \\
-{12\over m}c_5v^\mu\d_\mu\Phi\Bigl\}
\label{3.17}
\eea
and
\bea
N^\mu=8(u^\mu h^{\al\beta}-u^\beta h^{\al\mu})\d_\al\Phi\d_\beta\Phi
-(2M^2+2mc_3)u^\mu h\Phi^2
\nonumber \\
-2c_3(2u^\mu\d_\al v^\al-u^\al\d_\al v^\mu)\Phi^2.
\label{3.18}
\eea
\end{thm}
{\bf Proof:}

In this proof we must calculate all anomalies containing 
$\Phi$. 
We also give the commutators where the anomalies come from. From
\bea
(-u^\al\d_\al h-\d_\al u^\al h+2\d^\nu u^\al h_{\al\nu})[\d_\mu h(x),h(y)]
\left(mc_3-{M^2\over 2}c_4\right)\Phi^2
\nonumber
\eea
we get the anomaly
\be
A_1=2i(2mc_3-M^2c_4)(-u^\al\d_\al h-\d_\al u^\al h+2\d^\nu u^\al h_{\al\nu})
\Phi^2\delta,
\ee
and
\bea
2(u^\la\d_\la h^{\al\nu}+\d_\la u^\la h^{\al\nu}-\d_\la u^\al h^{\la\nu}
-\d_\la u^\nu h^{\al\la})[\d_\mu h_{\al\nu}(x),h(y)]
\left(mc_3-{M^2\over 2}c_4\right)\Phi^2
\nonumber
\eea
leads to
\be
A_2=i(2mc_3-M^2c_4)(u^\la\d_\la h+\d_\la u^\la h-2\d_\la u^\al h_{\al\la})
\Phi^2\delta,
\ee
The commutator
\bea
(-u^\al\d_\al h-\d_\al u^\al h+2\d^\nu u^\al h_{\al\nu})[\d_\mu h(x),
h^{\beta\gamma}(y)]c_4\d_\beta\Phi\d_\gamma\Phi
\nonumber
\eea
gives
\be
A_{3} = ic_4(-u^\al\d_\al h-\d_\al u^\al h+2\d^\nu u^\al h_{\al\nu})\d_\beta\Phi
\d^\beta\Phi
\ee
and
\bea
2(u^\la\d_\la h^{\al\nu}+\d_\la u^\la h^{\al\nu}-\d_\la u^\al h^{\la\nu}
-\d_\la u^\nu h^{\al\la})[\d_\mu h_{\al\nu}(x),h^{\beta\gamma}(y)]
c_4\d_\beta\Phi\d_\gamma\Phi
\nonumber
\eea
yields
\bea
A_4=-ic_4[2u^\la\d_\la h^{\beta\gamma}+2\d_\la u^\la h^{\beta\gamma}-(u^\al 
\d_\la h+\d_\la u^\la h)\eta^{\beta\gamma}
\nonumber \\
-2\d_\la u^\beta h^{\gamma\la}-2\d_\la u^\gamma h^{\beta\gamma}+2\d_\la 
u_\al h^{\la\al}\eta^{\beta\gamma}]c_4\d_\beta\Phi\d_\gamma\Phi.
\eea

Next the commutator
\bea
-4u^\al\d_\al v^\nu[\d_\mu v_\nu(x),\d_\beta v^\beta(y)]2c_3\Phi^2
\nonumber
\eea
leads to
\be
A_5=-4ic_3u^\al\d_\al v^\nu(x)\Phi^2(y)\d_\nu^y\delta(x-y).
\ee
and finally
\bea
-c_4u^\beta\d_\beta\Phi\Bigl[\d_\mu\Phi(x),\Phi^2(y)\left(mc_3h+2c_3\d_\al v^\al 
-{c_4\over 2}M^2h\right)+c_4\d_\al\Phi(y)\d_\nu\Phi h^{\al\nu}+c_5\Phi^3(y)\Bigl]
\nonumber
\eea
gives
\be
A_{6} = ic_4u^\beta\d_\beta\Phi(x)\{2\Phi\left(mc_3h-{c_4\over 2}M^2\right)h+4c_3\Phi\d_\nu v^\nu
+2c_4\d_\al\Phi(y) h^{\al\nu}(y)\d_\nu^y+3c_5\Phi^2\}\delta(x-y).
\ee
The sum 
$
A_1+\ldots+A_6
$ 
is equal to
\bea
B_1=-2ic_4(u^\la\d_\la h^{\al\beta}+\d_\la u^\la h^{\al\beta}-\d_\la u^\beta
h^{\al\la}-\d_\la u^\al h^{\beta\la})\d_\al\Phi\d_\beta\Phi\delta\qquad 
\nonumber \\ 
+2ic_4^2u^\beta\d_\beta\Phi(x)h^{\mu\nu}(y)\d_\mu\Phi(y)\d_\nu^y\delta(x-y)
\quad (T1)\nonumber \\
-i(2mc_3-M^2c_4)(u^\mu\d_\mu h+\d_\mu u^\mu h)\Phi^2\delta\qquad
\nonumber \\
+ic_4(2mc_3-M^2c_4)u^\beta\d_\beta\Phi h\Phi\quad (T2)
\nonumber \\
+2i(2mc_3-M^2c_4)\d_\nu u_\mu h^{\mu\nu}\Phi^2\delta\quad (T3)
\nonumber \\
-4ic_3u^\mu\d_\mu v^\nu(x)\Phi^2(y)\d_\nu^y\delta(x-y)+4ic_3c_4u^\beta
\d_\beta\Phi\Phi\d_\mu v^\mu\delta\quad (T4)
\nonumber \\
+3ic_4c_5u^\beta\d_\beta\Phi\Phi^2\delta.\quad (T5)
\eea
Following the methods developed in \cite{S} (Sect.5.9) we have grouped the
terms according to their type of Lorentz contractions. 
For example, $(T1)$ has $u^\la h^{\al\beta}\Phi\Phi$ and 3 derivatives which is
different from $(T3)$.
Only the terms within one type 
$
T1,\ldots T4
$ 
can be combined to give a divergence. Due to the different coefficients 
$c_4$ and $c_4^2$ in T1 we must have $c_4=-2$ in order to get a divergence.
If $c_4$ were $\ne -2$ then the last term of $(T1)$ would remain without
compensation. Since this term is not a relative coboundary gauge invariance then
would be violated.

The total anomaly is obtained by adding the contribution 
$
x\leftrightarrow y
$
according to (\ref{3.1}). For the terms with 
$
\delta(x-y)
$ 
this simply gives  factor 2. For the terms with derivative of 
$\delta$ 
we use the identity
\be
g(x)f(y)\d_\al^y\delta(x-y)+x\leftrightarrow y=(\d_\al gf-g\d_\al f)\label{3.15}
\delta(x-y)
\ee

Now the total anomalies of type T1 in (3.14) can be written in the form
\bea
(T1)_{\rm tot}=-4ic_4\Bigl[(u^\la\d_\la h^{\al\beta}+\d_\la u^\la h^{\al\beta})
\d_\al\Phi\d_\beta\Phi-\d_\la u^\beta h^{\al\la}\d_\al\Phi\d_\beta\Phi
\nonumber \\
+u^\beta\d_\al\d_\beta\Phi\d_\la\Phi h^{\al\la}-u^\beta\d_\beta\Phi\d_\al
\Phi\d_\la h^{\al\la}-u^\beta\d_\beta\Phi\d_\al\d_\la\Phi h^{\al\la}.
\eea
This agrees with the result in massless gravity \cite{S}, eq.(5.9.40), and is a divergence
\bea
(T1)_{\rm tot}=-4ic_4\d_\la^x\Bigl[(u^\la h^{\al\beta}-u^\beta h^{\la\al})\d_\al\Phi\d_\beta
\Phi\delta(x-y)]
\nonumber \\
+ x\leftrightarrow y,
\eea
where 
$c_4=-2$ 
has been taken into account and will be assumed in the following. Type T2 is a divergence as well:
\be
(T2)_{\rm tot}=-2i(M^2+mc_3)\d_\mu[u^\mu h\Phi^2\delta(x-y)]+x\leftrightarrow y.
\ee

As in the massless case (\cite{S}, eq.(5.9.45)) T3 is a coboundary:
\be
(T3)_{\rm tot}=2(M^2+mc_3)d_Q[(h^2-2h_{\mu\nu}h^{\mu\nu})\Phi^2\delta(x-y)].
\ee
Using the identity (\ref{3.15}) we write T4 as follows
\bea
(T4)_{\rm tot}=-4ic_3[(\d_\nu u^\mu+u^\mu\d_\mu\d_\nu v^\nu)\Phi^2-2u^\mu\d_\mu v^\nu 
\Phi\d_\nu\Phi
\nonumber \\
+4u^\mu\d_\nu v^\nu\Phi\d_\mu\Phi]\delta(x-y).
\eea
We first split off a divergence
\bea
(T4)_{\rm tot}=-4ic_3[2\d_\mu (u^\mu\d_\nu v^\nu\Phi^2)-\d_\nu (u^\mu\d_\mu v^\nu\Phi^2)
\nonumber \\
+2(\d_\nu u^\mu\d_\mu v^\nu-\d_\mu u^\mu\d_\nu v^\nu)\Phi^2]\delta(x-y).
\eea
Now the terms in the second line are a coboundary
\bea
(T4)_{\rm tot}=-2ic_3[2\d_\mu^x (u^\mu\d_\nu v^\nu\Phi^2\delta)-\d_\nu^x 
u^\mu\d_\mu v^\nu\Phi^2\delta)]+x\leftrightarrow y
\nonumber \\
+8{c_3\over m}d_Q[(\d_\nu v^\mu\d_\mu v^\nu-\d_\mu v^\mu\d_\nu v^\nu)\Phi^2\delta].
\eea
Finally, T5 is a coboundary
\be
T5={12\over m}c_5d_Q(v^\mu\Phi^2\d_\mu\Phi\delta).
\ee
Adding the contribution 
$
x\leftrightarrow y
$ 
this gives the result of the theorem.
$\qed$

\section{Modified general relativity}

As in ref. \cite{Sch1} we now consider the limit 
$
m\to 0
$ of vanishing graviton mass. The point is that this does not lead to massless gravity because the vector graviton field 
$
v^\mu
$ does not decouple from the other fields. In fact, in first order (proportional to Newton's constant) there survives the coupling term
\be
T_v=h^{\mu\nu}\d_\mu v^\la\d_\nu v_\la.
\ee
If scalar matter is included then in addition to the standard coupling to the energy-momentum tensor of the scalar field ($\sim c_4$ in (\ref{3.13})) two further couplings 
$
T_3
$ 
and 
$
T_5
$ 
are possible. However, in second order the graviton mass appears in the denominator in $N$ in (\ref{3.17}). Consequently, if the coupling constants 
$
c_3
$ 
and 
$
c_5
$ 
do not depend on $m$, the limit
$
m\to 0
$ 
exists for 
$
c_3=0=c_5
$, 
only. Then we have no direct coupling of the $v$-field to normal matter; the resulting theory of \cite{Sch1} seems not to be physically relevant.

There is another option. Gauge invariance does not forbid the possibility that 
$
c_3
$ and 
$
c_5
$ depend on $m$, for example
\be
c_3=\la_3 m,\quad c_5=\la_5m,
\ee
where 
$
\la_j
$ 
are independent of $m$. Then in the limit 
$
m\to 0
$ 
the first order trilinear couplings die away, but there remain the following quartic couplings from second order 
\be
T_{\Phi v}=2\Phi^2\Bigl\{8\la_3(\d_\mu v^\mu\d_\nu v^\nu-\d_\mu v^\nu\d_\nu v^\mu)
-12\la_5v^\mu\d_\mu\Phi\Bigl\}.
\label{4.3}
\ee
In the classical limit this coupling must be added to the classical Lagrangean. As in the
other coupling terms the usual factor 
$
\sqrt{-g}
$ 
is included. The necessity of this factor becomes clear when we derive the field equations below; but of course, an independent check by a third order calculation must be done. Our modification of general relativity is now defined by the following Lagrangean density
\bea
L_{\rm tot}={-2\over\kappa^2}\sqrt{-g}R+\sqrt{-g}g^{\mu\nu}\d_\mu v_\la\d_\nu v^\la
\nonumber \\
+{1\over 2}\sqrt{-g}g^{\mu\nu}\d_\mu\Phi\d_\nu\Phi+\sqrt{-g}\Phi^2\Bigl\{\la_3(\d_\mu v^\mu\d_\nu v^\nu-\d_\mu v^\nu\d_\nu v^\mu)+\la_5v^\mu\d_\mu\Phi\Bigl\}.
\label{4.4}
\eea
The two terms in the first line are the pure gravitational interactions which have been
studied already in \cite{Sch1}. The first term is the Einstein-Hilbert Lagrangean, 
$
\kappa^2=32\pi G
$ 
is essentially Newton's constant and $R$ the scalar curvature. The second line contains the interaction with scalar matter; the numerical factors in (\ref{4.3})  have been absorbed by redefining the coupling constants
$
\la_3
$ 
and 
$
\la_5
$. 

The Lagrangean (\ref{4.4}) as it stands is Lorentz invariant, but the new terms in the second line are not invariant under general coordinate transformations. In \cite{Sch1} we have argued that this latter invariance can be maintained in the second term of the first
line, if we consider 
$
v^\la
$ as four scalar fields. This argument cannot be used for the new matter couplings in the second line. The lack of general covariance might be disturbing for classical relativists. However, one should keep in mind that classical general covariance corresponds to gauge invariance of the spin-2 quantum gauge theory, so that this principle is incorporated in the quantum theory. The latter is background dependent; we have selected Minkowski background. Returning again to the classical theory this background dependence remains and we get a Lorentz invariant classical theory, not a general covariant one. Still, by checking gauge invariance in third order we have to test whether there are further modifications in the classical theory. This will be done elsewhere.

The Euler-Lagrange equations for the Lagrangean (\ref{4.4}) give the system of coupled
field equations. Variation of 
$
g^{\mu\nu}
$ 
gives the modified Einstein equations
\bea
R_{\mu\nu}-{1\over 2}g_{\mu\nu}R={16\pi G\over c^3}\Bigl\{\d_\mu v_\la\d_\nu v^\la
-{1\over 2}g_{\mu\nu}g^{\al\beta}\d_\al v_\la\d_\beta v^\la
\nonumber \\
 +{1\over 2}\d_\mu\Phi
\d_\nu\Phi-{1\over 4}g_{\mu\nu}(g^{\al\beta}\d_\al\Phi\d_\beta\Phi-M^2\Phi^2)
\nonumber \\
-{1\over 2}g_{\mu\nu}\Phi^2\Bigl[\la_3(\d_\al v^\beta\d_\beta v^\al-\d_\al v^\al 
\d_\beta v^\beta)+\la_5 v^\al\d_\al\Phi\Bigl]\Bigl\}.
\eea
The variational derivative with respect to 
$
v^\mu
$ 
yields
\bea
\d_\al(\sqrt{-g}g^{\al\beta}\d_\beta v_\mu)=2\la_3\Bigl[\d_\mu(\sqrt{-g}\Phi^2\d_\nu 
v^\nu)-\d_\nu(\sqrt{-g}\Phi^2\d_\mu v^\nu)\Bigl]
\nonumber \\
+\la_5\sqrt{-g}\Phi^2\d_\mu\Phi.
\eea
Here the vector-graviton field has source terms from the new scalar-matter coupling.
Note that the second order derivative 
$
\d_\mu\d_\nu v^\nu
$ 
cancels on the right-hand side so that we have a wave equation with source. Finally, the variation of 
$\Phi$ 
gives the Klein-Gordon equation in the metric 
$
g^{\al\beta}
$ 
plus source terms:
\bea
{1\over\sqrt{-g}}\d_\al(\sqrt{-g}g^{\al\beta}\d_\beta \Phi)+M^2\Phi= -2\la_3\Phi(\d_\mu v^\mu  \d_\nu v^\nu)-\_\nu v^\mu\_\mu v^\nu)
\nonumber \\
-\la_5\Phi^2{1\over\sqrt{-g}}\d_\mu(\sqrt{-g}v^\mu).
\eea
The physical consequences of these field equations remain to be investigated,
in particular whether there are solutions giving an explanation of the dark matter
phenomenology.

\end{document}